\documentstyle[epsfig, 12pt]{article}
\begin{document}

\begin{titlepage}

\hfill{TMUP-HEL-97/07}

\hfill{UM-P-97/50}

\hfill{RCHEP-97/10}

\vskip 2.0 cm

\centerline{{\large \bf 
Testing maximal electron and muon neutrino oscillations}}
\centerline{{\large \bf
with sub-GeV SuperKamiokande atmospheric neutrino data }}
\vskip 1.0 cm
\centerline{R. Foot$^a$, R. R. Volkas$^a$ and O. Yasuda$^b$}
\vskip 0.7 cm
\noindent
\centerline{{\it $^a$ School of Physics,}}
\centerline{{\it Research Centre for High Energy Physics,}}
\centerline{{\it The University of Melbourne,}}
\centerline{{\it Parkville, 3052 Australia. }}
\vskip 0.5cm
\noindent
\centerline{{\it $^b$ Department of Physics, 
Tokyo Metropolitan University,}}
\centerline{{\it 1-1 Minami-Osawa Hachioji, Tokyo
192-03, Japan}}

\vskip 1.0cm

\centerline{Abstract}
\vskip 0.7cm
\noindent

Motivated by the Exact Parity Model and other theories, the hypothesis that
each of the known neutrinos oscillates maximally with a sterile partner has
been put forward as an explanation of the atmospheric and solar neutrino
anomalies.  We provide detailed predictions for muon and electron flux ratios
induced in the Kamiokande and SuperKamiokande detectors by sub-GeV
atmospheric neutrinos. Several different, carefully chosen cuts on momentum and
zenith angle are proposed, 
emphasizing the role of up-down flux asymmetries.

\end{titlepage} 

The solar \cite{solar} and atmospheric \cite{kamioka1,atmospheric}
neutrino observations and the LSND \cite{lsnd} experiment provide strong
evidence that neutrinos have nonzero masses and oscillate. Many specific
neutrino mass scenarios have been proposed as explanations for some or all of
these results. In this paper we will focus on the hypothesis that each of
the known neutrinos oscillates maximally with an 
exactly or effectively sterile
partner. Though non-minimal, this scenario is theoretically 
well motivated by
the Exact Parity Model \cite{flv} and other theories \cite{oth}. 

Denote the three sterile neutrinos by $\nu'_e$, $\nu'_{\mu}$, and
$\nu'_{\tau}$. In the Exact Parity Model, all known particles are paired via
an exact parity symmetry with mirror partners in order to achieve a parity
symmetric dynamics. Since parity eigenstates must also be energy eigenstates,
the mass eigenstate neutrinos are maximal mixtures of $\nu_{\alpha}$ and
$\nu'_{\alpha}$ (where $\alpha = e,\mu,\tau$) in the limit of small
intergenerational mixing. The models of Ref.\cite{oth} provide approximate
maximal mixing via an approximate discrete symmetry of the neutrino mass
sector. We will assume small intergenerational mixing, analogous to the quark
sector, for the purposes of this paper.

Maximal $\nu_e - \nu'_e$ and $\nu_{\mu} - \nu'_{\mu}$ oscillations have been
put forward as explanations of the solar and atmospheric neutrino anomalies
respectively \cite{flv,oth}.  This scenario is also compatible with the LSND
experiment.  Maximal $\nu_e - \nu'_e$ oscillations can solve the solar
neutrino problem for $\delta m^2_{ee'}$ in the large range\cite{gs,js},
\begin{equation} 
3 \times 10^{-10} \stackrel{<}{\sim} |\delta m^2_{ee'}|/eV^2
\stackrel{<}{\sim} 0.9 \times 10^{-3}, 
\label{1} 
\end{equation} 
where the
upper bound is the most recent experimental limit \cite{exp}.  Maximal
$\nu_{\mu} - \nu'_{\mu}$ oscillations can explain the atmospheric neutrino
anomaly \cite{cos} provided that \cite{kamioka1,barpak}
\begin{equation} 
10^{-3} \stackrel{<}{\sim} |\delta m^2_{\mu\mu'}|/eV^2 
\stackrel{<}{\sim} 10^{-1}.
\end{equation}  
Note however that the atmospheric neutrino experiments are sensitive to both
$\nu_{\mu} -\nu'_{\mu}$ and $\nu_e - \nu'_e$ oscillations in principle. The
implications of this for atmospheric neutrino experiments were
discussed qualitatively in Ref.\cite{bfv}, where rather crude analytic
estimates were provided of some relevant measurable quantities. 
The purpose of this paper is to provide a detailed quantitative study of the
implications of maximal $\nu_\mu - \nu'_\mu$ and $\nu_e - \nu'_e$
oscillations for the Kamiokande and SuperKamiokande experiments.  

We will restrict our analysis to sub-GeV neutrinos. The main reason for this
is as follows. Consider the zenith-angle averaged ``ratio of ratios''
$\langle R \rangle$, where 
\begin{equation} 
R \equiv \frac{(N_{\mu}/N_e)_{obs}}{(N_{\mu}/N_e)_{MC}}.  
\end{equation}
The quantities $N_{e,\mu}$ are the numbers of electron- and 
muon-like events. 
The numerator denotes numbers actually observed, while the denominator 
the numbers expected on the basis of a Monte-Carlo simulation
without oscillations.  Since $\langle R \rangle$ is measured to be
significantly less than 1, we conclude that the $\nu_e$ oscillation length
must be much larger than the $\nu_{\mu}$ oscillation length.  Given this,
$\nu_e$ oscillations are expected to impact more significantly on the sub-GeV
data than the multi-GeV data, because the neutrino oscillation length
decreases with energy. We will not comment
further on the multi-GeV neutrinos, except to note that the Kamiokande and
especially the preliminary SuperKamiokande data \cite{SKprelim} are
consistent with maximal $\nu_{\mu} - \nu'_{\mu}$ oscillations and are
inconsistent with the minimal standard model. 

In the water-Cerenkov Kamiokande and SuperKamiokande experiments,
neutrinos are detected via the charged leptons $\ell_\alpha$~
($\ell_\alpha$ = $e$ or $\mu$) produced primarily from quasi-elastic
neutrino scattering off nucleons in
the water molecules: $\nu_\alpha N \rightarrow \ell_\alpha N'$
$(\alpha=e,\mu)$. 
The total number $N(\ell_\alpha)$ of charged leptons of
type $\ell_{\alpha}$ produced through $\nu_\alpha N \rightarrow \ell_\alpha
N'$ is given by
\begin{eqnarray}
\displaystyle
N(\ell_\alpha)
&=& n_T \
\int_0^\infty dE
\int^{q_{\rm max}}_{q_{\rm min}} dq
\int_{-1}^1 d\cos \psi
\int_{-1}^{+1} d\cos \theta \nonumber\\
&\times&
{d^2F_\alpha (E,\theta) \over dE~d\cos\theta}
\cdot{ d^2\sigma_\alpha (E,q,\cos\psi) \over dq~d\cos\psi }
\cdot
{\ }P(\nu_\alpha\rightarrow\nu_\alpha; E, \theta).
\label{eqn:n}
\end{eqnarray}
Here $d^2F_\alpha /dEd\cos\theta$ is the differential flux of atmospheric
neutrinos of type $\nu_\alpha$ of energy $E$ 
at zenith angle $\theta$.
The term $n_T$ is the effective number of target nucleons.
The function $d^2\sigma_\alpha/dqd\cos\psi$ is the differential cross section
for quasi-elastic scattering,
$\nu_\alpha N \rightarrow \ell_\alpha N'$, where $\psi$ is the scattering
angle relative to the velocity vector of the incident $\nu_{\alpha}$ (the
azimuthal angle having been integrated over), and $q$ is the energy of the
charged lepton $\ell_{\alpha}$. The function
$P(\nu_\alpha\rightarrow\nu_\alpha; E, \theta)$ is the survival
probability for a $\nu_\alpha$ with energy $E$ after
traveling a distance
$\displaystyle L=\sqrt{(R+h)^2-R^2\sin^2\theta}-R\cos\theta$,
where $R$ is the radius of the Earth and $h\sim$ 15 km is the altitude
at which atmospheric neutrinos are produced.
It can be obtained by solving the
Schr\"odinger equation for neutrino evolution including matter effects (for a
review see, for example, Ref.\cite{gr}).
In our analysis we have a pair of two-flavour oscillation subsystems,
$\nu_e - \nu'_e$ and $\nu_{\mu} - \nu'_{\mu}$, where
the vacuum mixing angles are maximal ($\sin^2 2\theta_0^{ee'} = 1$ and
$\sin^2 2\theta_0^{\mu\mu'} = 1$).
There are thus only two free parameters: $\delta m^2_{ee'}$
and $\delta m^2_{\mu \mu'}$. Note that Eq.(\ref{eqn:n}) 
must be modified in order
to obtain zenith angle and momentum binned charged lepton events.

For the case of sub-GeV neutrinos, almost all the information
necessary for the right hand side of Eq.(\ref{eqn:n}) is
available in the published references.  We have used the differential
cross section $d^2\sigma_\alpha / dqd\cos\psi$ of Ref.\cite{og}.
The differential flux of atmospheric neutrinos
$d^2F_\alpha /dEd\cos\theta$ without geomagnetic effects
is given in \cite{hkkm}, but we have used the differential
flux which includes geomagnetic effects \cite{hkm}.

For reference and for completeness, we first briefly discuss the case where
only maximal $\nu_{\mu} - \nu'_{\mu}$ oscillations occur.
Figure 1 shows the
variation of $\langle R \rangle$ for sub-GeV neutrinos with $\delta m^2_{\mu
\mu'}$. The preliminary SuperKamiokande central value plus the $1\sigma$
confidence level band is also plotted. The datum implies $10^{-3}
\stackrel{<}{\sim} \delta m^2_{\mu\mu'}/eV^2 \stackrel{<}{\sim} 10^{-2}$ at
the $1\sigma$ level, with a much larger range allowed at the $2\sigma$ level
(but also constrained by the zenith angle dependence of the multi-GeV
sample). Note that while a major thrust of this paper is to examine possible
manifestations of $\nu_e - \nu'_e$ oscillations, it is quite possible that
$\delta m^2_{ee'}$ is too small to lead to any observable effects for
atmospheric neutrinos. In that case, $\nu_{\mu} - \nu'_{\mu}$ oscillations
will be the only effect occurring. This case is almost indistinguishable from
the oft-considered maximal $\nu_{\mu} - \nu_{\tau}$ oscillation solution to
the atmospheric neutrino anomaly\cite{barpak}. Note that future long baseline
$\nu_{\tau}$-appearance experiments should help distinguish the two cases. 

We now turn on maximal $\nu_e - \nu'_e$ oscillations.
As discussed qualitatively in Ref.\cite{bfv}, there
are several interesting effects that can 
occur when both $\nu_e - \nu'_e$
and $\nu_{\mu} - \nu'_{\mu}$ oscillations are present.
One simple effect of $\nu_e - \nu'_e$ oscillations, for a given 
$\delta m^2_{\mu \mu'}$,
is to increase $\langle R \rangle$ relative to what it would be in the
absence of these oscillations.
We illustrate this in Fig.\ 2, which gives 
$\langle R \rangle$ as a function of $\delta m^2_{ee'}$
for various values of $\delta m^2_{\mu \mu'}$.
We have used the usual Kamiokande momentum cuts
$0.1 < p_e/GeV < 1.33$ and $0.2 < p_{\mu}/GeV < 1.5$.
The current experimental measurements of $\langle R \rangle$
by the Kamiokande and SuperKamiokande collaborations are \cite{SKprelim},
\begin{eqnarray}
& \langle R \rangle_{Kam} = 0.60^{+0.06}_{-0.05}\pm 0.05&  
\nonumber \\
& \langle R \rangle_{SKam} = 0.635\pm 0.033\pm 0.053 
\ ({\rm preliminary}).& 
\end{eqnarray}
In the absence of $\nu_e - \nu'_e$ oscillations (or
equivalently, when $|\delta m^2_{ee'}| \stackrel{<}{\sim}
2\times 10^{-5}\ eV^2$) the above measurements suggest
$|\delta m^2_{\mu \mu'}| \sim 5 \times 10^{-3}\ eV^2$.
However, as Fig.\ 2 shows, much larger values of 
$\delta m^2_{\mu \mu'}$ can fit these data equally well when
$\nu_e - \nu'_e$ oscillations also occur. For example, an $\langle R \rangle$
value of about $0.6$ is obtained for $\delta m^2_{\mu\mu'} \sim 6
\times 10^{-2}\ eV^2$ when $\delta m^2_{ee'} \sim 10^{-4}\ eV^2$.
Figure 2 also shows that all of the range of Eq.(\ref{1})
is consistent with the (Super)Kamiokande data.

The observation that $\nu_e - \nu'_e$ oscillations increase $\langle R
\rangle$ is interesting for future long baseline neutrino oscillation
experiments. If $\nu_e - \nu'_e$ oscillations are numerically important for
atmospheric neutrinos, then the long baseline determination of $\delta
m^2_{\mu\mu'}$ should not be consistent with that implied by atmospheric
neutrinos under the assumption that $\nu_{\mu} - \nu'_{\mu}$ oscillations
only are occurring. A discrepancy would signal that an additional effect is at
work for atmospheric neutrinos, with $\nu_e - \nu'_e$ oscillations being a
candidate. 

One way to experimentally uncover the presence of both $\nu_{\mu} -
\nu'_{\mu}$ and $\nu_e - \nu'_e$ oscillations is through the energy
dependence of $\langle R \rangle$. It turns out that a more
sensitive way is through the zenith-angle dependence 
of fluxes \cite{bfv}, as we now explain. 

Consider the quantities $Y_e^0$ and $Y_{\mu}^0$ where
\begin{equation}
Y^0_e \equiv {(N_e^-/N_e^+)|_{osc} \over (N_e^-/N_e^+)|_{no-osc}},\qquad
Y^0_\mu \equiv {(N_\mu^-/N_\mu^+)|_{osc} \over (N_\mu^-/N_\mu^+)|_{no-osc}}.
\label{Y}
\end{equation}
Here $N_e^-$ ($N_e^+$) is the number of electrons produced in
the detector with zenith angle $\cos \Theta < 0$ ($\cos \Theta > 0$).
$N_{\mu}^{\pm}$ are the analogous quantities for muons. (The zenith
angle $\Theta$ for the charged leptons should not be confused with the zenith
angle $\theta$ defined earlier for neutrinos.) The numerators are the
predictions for $N_{\alpha}^-/N_{\alpha}^+$ $(\alpha = e, \mu)$ in the model
while the denominators are the same quantities in
the absence of oscillations. (Note that
$N_{\alpha}^-/N_{\alpha}^+|_{no-osc}$
are close to 1 for symmetry reasons, but not exactly equal to it.)  
The numerators and denominators,
being ratios of fluxes, should be approximately free
of the systematic errors arising from uncertain absolute fluxes ($\pm
30\%$), uncertain $\mu$ to $e$ flux ratios ($\pm$ a few percent) and the
uncertain cross-section. The
$Y^0_{\alpha}$ are a 
measure of up-down flux asymmetries. It is important to observe that $Y^0_e$ 
depends only on $\delta m^2_{e e'}$, and $Y_\mu$ 
depends only on $\delta m^2_{\mu \mu'}$.

We have plotted $Y^0_e$ as a function of $\delta m^2_{ee'}$ 
in Fig.\ 3 \cite{fn}. 
We have used the usual Kamiokande cut $0.1 < p_e/GeV < 1.33$ as well an the
alternative $0.5 < p_e/GeV < 1.33$ cut. In the former case we see that
$Y^0_e$ decreases from 1 to about $0.82$ as $\delta m^2_{ee'}$ is
increased from $10^{-5}\ eV^2$ towards the experimental limit. The effect is
much more pronounced when the alternative cut on momentum is made, with
$Y^0_e$ decreasing to about $0.67$. The reason for this is the greater
correlation between the zenith angle distribution of the electrons and the
zenith angle distribution of the incident $\nu_e$ flux when the average
energy of the sample is increased. A price has to be paid in statistics since
the neutrino flux is larger at smaller energies, but a gain is made in the
size of the signal relative to the underlying systematic error (which is at
the few percent level for flux ratios). This is important because the
statistical error will be reduced as the data sample grows, whereas the
systematic error will remain the same (barring great improvements in the
theoretical prediction of neutrino fluxes). 

Figures 4 and 5 are similar to Fig.\ 3, except we have redefined $N_e^+$ to
be the number of electrons with $\cos\Theta$ greater than $0.2$ and $0.6$,
respectively. Similarly $N_e^-$ is redefined to be the number of electrons
with $\cos\Theta$ less than $-0.2$ and $-0.6$, respectively. We use the
notation $Y^{0.2}_e$ and $Y^{0.6}_e$ for the quantities corresponding to
Eq.\ref{Y}. Note that these cuts on the zenith angle also increase the
magnitude of the effect, again with a concomitant decrease in statistics but
a gain in signal relative to systematic error.

Figures 6-8 are the corresponding figures for the muon case,
$Y_{\mu}$. Note that in this case the usual Kamiokande momentum cut is
$0.2 < p_{\mu}/GeV < 1.5$ and we have considered the 
alternative cut of $0.5 <
p_{\mu}/GeV < 1.5$ also. The peak at high $\delta m^2_{\mu \mu'}$ is due to
an oscillation node in the downward-going neutrinos. For the muon case there
is already interesting evidence for a zenith angle dependence of the flux of
multi-GeV neutrinos. In this context, we would like to emphasize the
important role that the ratios $Y_{\mu}$ and $Y_e$, with the
various momentum and zenith angle cuts, could play in determining whether the
atmospheric neutrino problem is due to $\nu_{\mu}$ oscillations alone or to
both $\nu_{\mu}$ and $\nu_e$ oscillations. Current data clearly show that
$\nu_{\mu}$ oscillations are the dominant effect.  However, the presence of
non-trivial $\nu_e$ oscillations cannot yet be ruled out and will be further
tested as SuperKamiokande continues to acquire data.

Note the insensitivity of $Y_{\mu}$ to
$\delta m^2_{\mu \mu'}$ for the quite large range
$3\times 10^{-4} \stackrel{<}{\sim}
\delta m^2_{\mu \mu'}/eV^2 \stackrel{<}{\sim} 10^{-2}$. 
The corresponding values of about $0.75-0.80$ and $0.65-0.70$ for
$Y^0_{\mu}$ are, in a sense, the most ``likely'' ones.
This insensitivity may provide
a useful test of maximal $\nu_{\mu} - \nu'_{\mu}$ oscillations and may also
be useful in discriminating against other possible explanations of
the atmospheric neutrino anomaly.

We will now quantify the decrease in statistical error expected with time.
After about 1 year of running, the preliminary SuperKamiokande data show
statistical errors for $Y_{e,\mu}$ of about $8\%$ and $12\%$ for the
standard
momentum cut and alternative momentum cut lower limit of $0.5\ GeV$,
respectively. These should decrease to about $3-4\%$ and $5-7\%$,
respectively, after 5 years of running. These are small enough to see or 
rule out many of the effects plotted in Figs.\ 3-8. 

In conclusion, we have provided a quantitative analysis of the
implications of the hypothesis that both the $\nu_{\mu}$ and $\nu_e$
neutrinos are maximally mixed with sterile partners for the
(Super)Kamiokande atmospheric neutrino experiments.
Maximal mixing,
motivated by the Exact Parity Model \cite{flv} and other theories \cite{oth},
implies that the oscillations can be described by just two parameters,
$\delta m^2_{ee'}$ and $\delta m^2_{\mu \mu'}$. As Figs.\ 2-5 show, if
$|\delta m^2_{ee'}| \stackrel{<}{\sim} 2\times
10^{-5}\ eV^2$, then the $\nu_e - \nu'_e$ oscillation length is too long
to affect the experiments.  As is well known, the experimental
results can then be explained by maximal $\nu_{\mu} - \nu'_{\mu}$
oscillations with $\delta m^2_{\mu \mu'}$ in a range around $5 \times
10^{-3}\ eV^2$.  Zenith angle asymmetries for
the detected sub-GeV muons produced are then expected (see Figs.\ 6-8).
(Significant zenith angle
dependence for multi-GeV muons has, of course, already been observed.)  For
$|\delta m^2_{ee'}| \stackrel{>}{\sim} 2 \times 10^{-5} \ eV^2$, the effects
of $\nu_e - \nu'_e$ oscillations become significant.  They have two main
effects. First, they can increase $\langle R \rangle$ (see Fig.\ 2)
relative to what it would be in their absence. Second, they can lead to
zenith angle asymmetries for the electrons produced in the detector (see
Figs.\ 3-5).  While the current data are consistent with the minimal
case of maximal $\nu_{\mu} -\nu'_{\mu}$ oscillations only, the possibility
that the electron neutrino also oscillates maximally with $|\delta m^2_{ee'}|
\stackrel{>}{\sim} 2 \times 10^{-5} \ eV^2$ is an interesting possibility.
Furthermore, this possibility will be tested stringently in the
near future as more data is collected, especially at SuperKamiokande.  

\vspace{5mm}
 
\noindent
{\large \bf Note added}

\vspace{5mm}

While this paper was in the final stages of preparation,
a paper by J. W. Flanagan, J. G. Learned
and S. Pakvasa appeared on the hep-ph archive (9709438), 
which also discusses the importance of up-down zenith 
angle asymmetries of the charged leptons induced by
atmospheric neutrinos.

\vspace{5mm}

\noindent 
{\large \bf Acknowledgements} 

\vspace{5mm} 

\noindent
R.F. would
like to acknowledge the hospitality of the Tokyo Metropolitan University,
where this work was initiated.  This work was partially supported by the
Australian Research Council. O.Y. was supported in part by a Grant-in-Aid
for Scientific Research of the Ministry of Education, Science and
Culture, \#09045036.

\newpage

\centerline{\large \bf Figure Captions}

\noindent
Figure 1.\ \ The zenith angled averaged ``ratio of ratios'', denoted $\langle
R \rangle$, plotted as a function of $\delta m^2_{\mu\mu'}$ for the case
where $\nu_e - \nu'_e$ oscillations are unimportant ($\delta m^2_{ee'}
\stackrel{<}{\sim} 2 \times 10^{-5} \ eV^2$). The usual
SuperKamiokande momentum cuts have been assumed. The band denotes the
$1\sigma$ allowed region from preliminary SuperKamiokande data.

\vspace{5mm}

\noindent
Figure 2.\ \ $\langle R\rangle$ plotted as a function of $\delta m^2_{ee'}$
for several values of $\delta m^2_{\mu\mu'}$ with the usual momentum cut used
by SuperKamiokande to define its sub-GeV sample.
The solid lines, going from top to bottom, correspond to 
$\delta m^2_{\mu \mu'}/eV^2  = 10^{-3}, 3\times 10^{-3}, 6\times
10^{-3}, 10^{-2}, 6\times 10^{-2}$ and $3\times 10^{-2}$, respectively. The
band denotes the $1\sigma$ allowed region from preliminary SuperKamiokande
data.

\vspace{5mm}

\noindent 
Figure 3.\ \ The quantity $Y^0_e$, defined in the text, as a
function of $\delta m^2_{ee'}$. The solid line uses the usual 
cut $0.1 < p_e/GeV < 1.33$ on the
electron momentum $p_e$, while the dashed line uses the alternative cut $0.5 <
p_e/GeV < 1.33$.

\vspace{5mm}

\noindent
Figure 4.\ \ Similar to Fig.\ 3 except we have plotted $Y_e^{0.2}$ defined
in the text.

\vspace{5mm}

\noindent
Figure 5.\ \ Similar to Fig.\ 3 except we have plotted $Y_e^{0.6}$ defined
in the text. 

\vspace{5mm}

\noindent 
Figure 6.\ \ The quantity $Y^{0}_{\mu}$, defined in the text, as a
function of $\delta m^2_{\mu\mu'}$. The solid line uses the 
usual cut $0.2 < p_{\mu}/GeV < 1.5$ on the
muon momentum $p_{\mu}$, while the dashed line uses the alternative cut $0.5 <
p_{\mu}/GeV < 1.5$.

\vspace{5mm}

\noindent 
Figure 7.\ \ Similar to Fig.\ 6 except we have plotted
$Y_{\mu}^{0.2}$ defined in the text.

\vspace{5mm}

\noindent
Figure 8.\ \ Similar to Fig.\ 6 except we have plotted $Y_{\mu}^{0.6}$
defined in the text. 

\pagestyle{empty}
\newpage
\epsfig{file=p1.eps,width=15cm}
\newpage
\epsfig{file=p2.eps,width=15cm}
\newpage
\epsfig{file=p3.eps,width=15cm}
\newpage
\epsfig{file=p4.eps,width=15cm}
\newpage
\epsfig{file=p5.eps,width=15cm}
\newpage
\epsfig{file=p6.eps,width=15cm}
\newpage
\epsfig{file=p7.eps,width=15cm}
\newpage
\epsfig{file=p8.eps,width=15cm}


\begin{thebibliography}{99}
\bibitem{fn}
In our numerical work we found some small oscillations of
the $Y$ parameter which we have smoothed out in Figs. 3-8
(the oscillations are typically less than 5 percent).
\end{thebibliography}
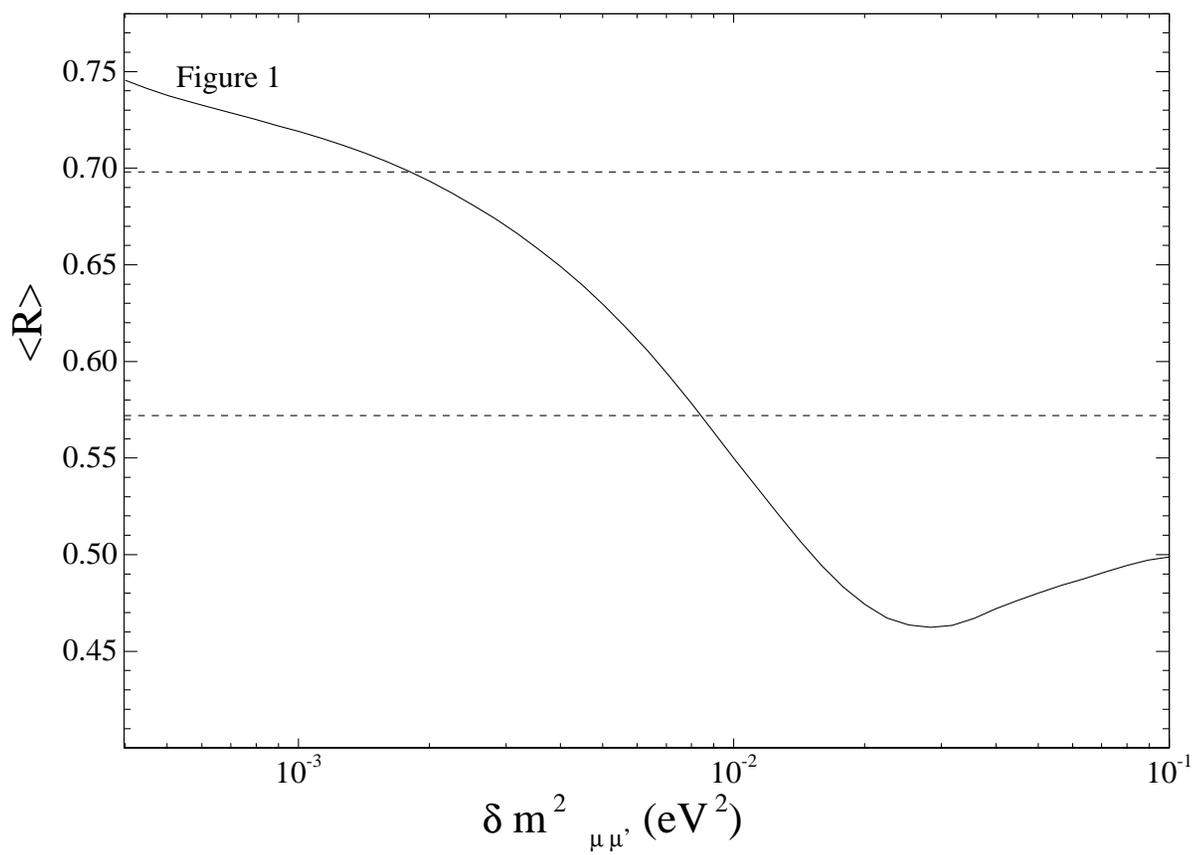
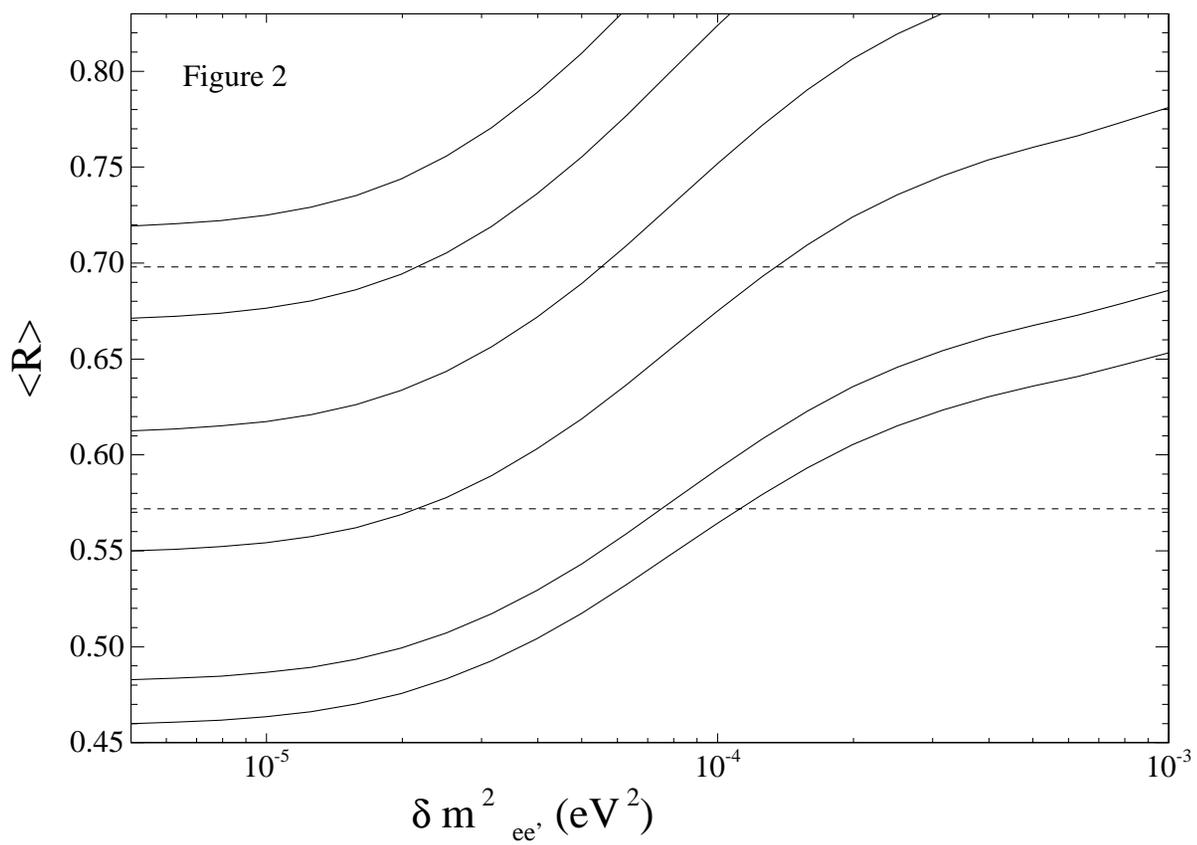
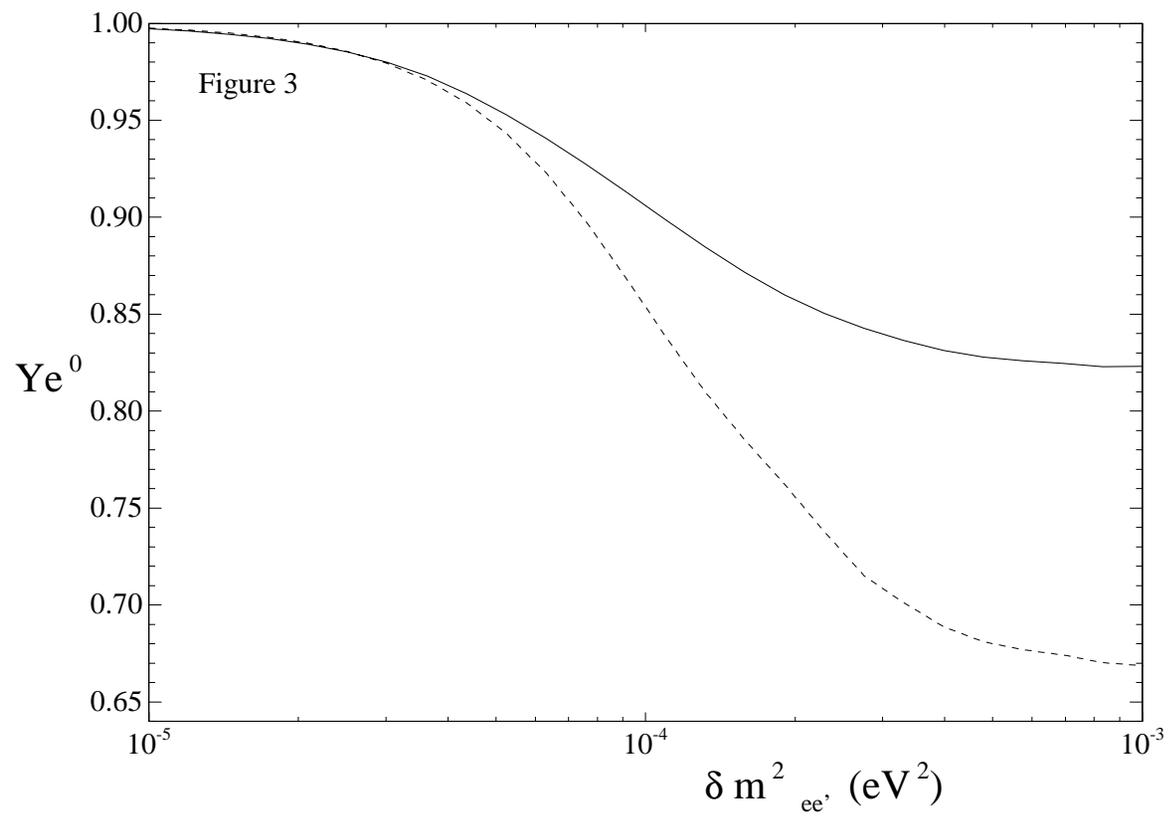
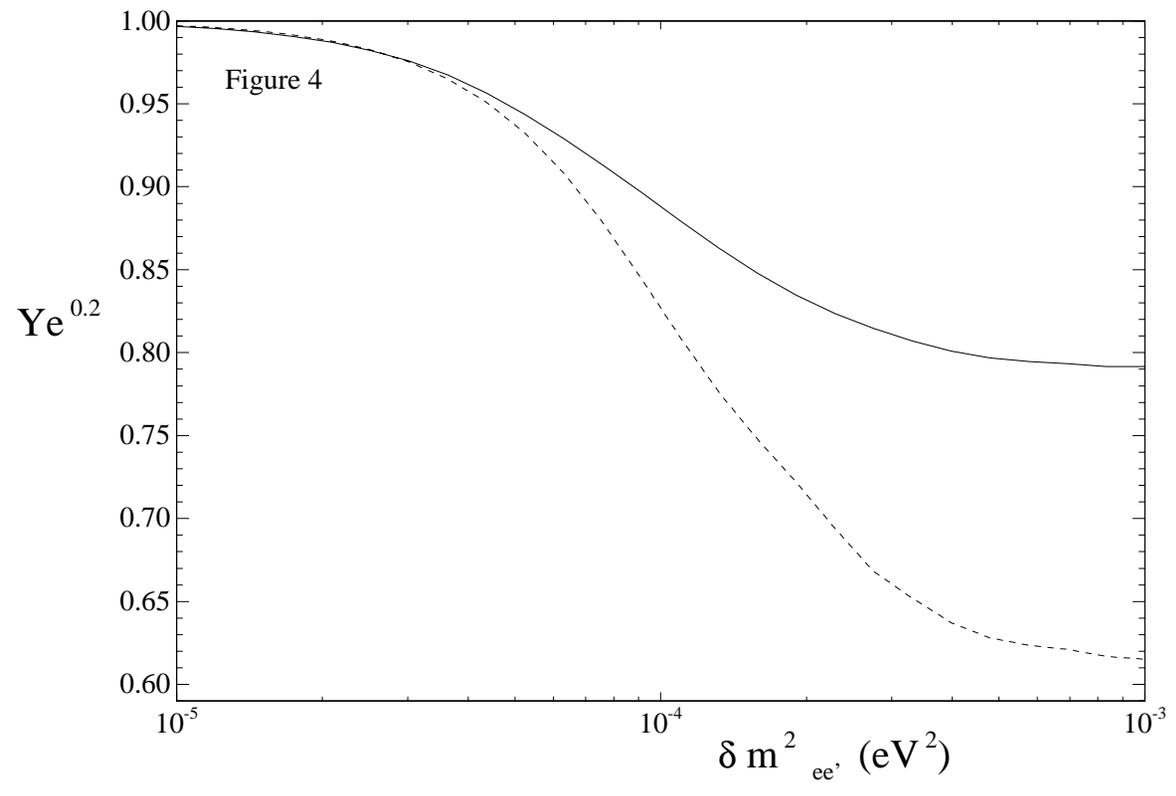
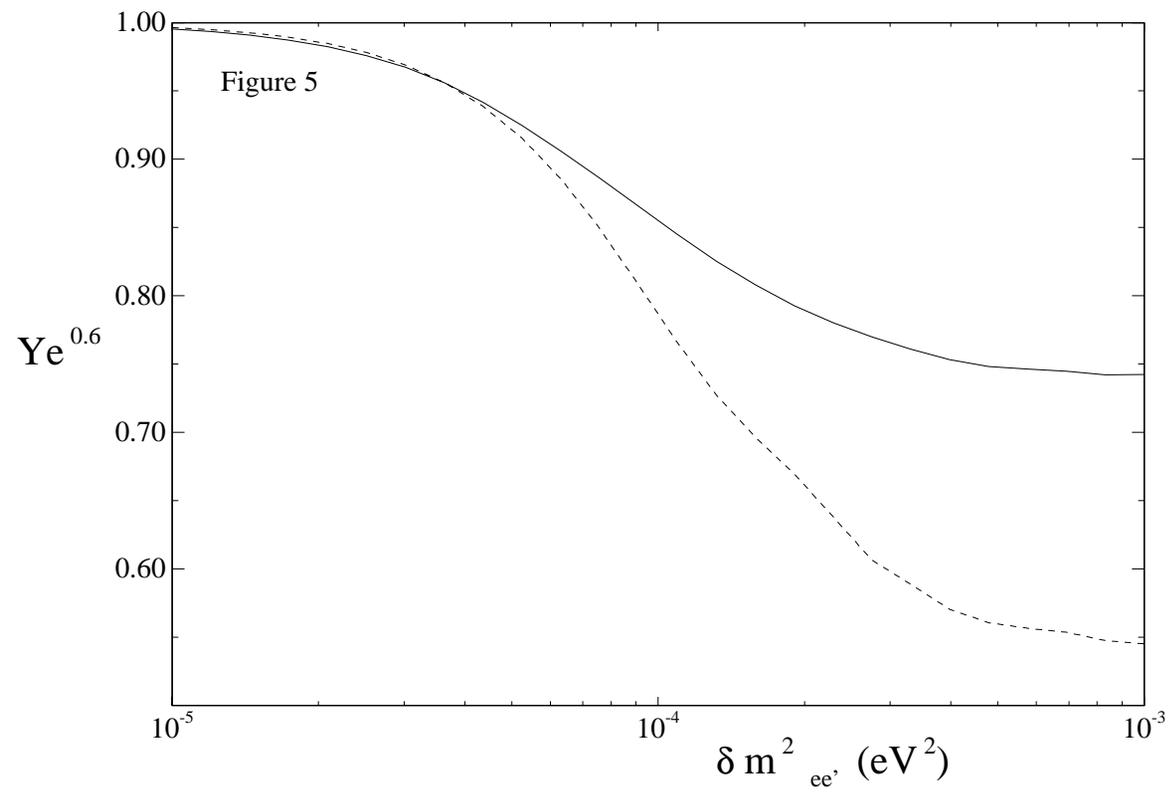
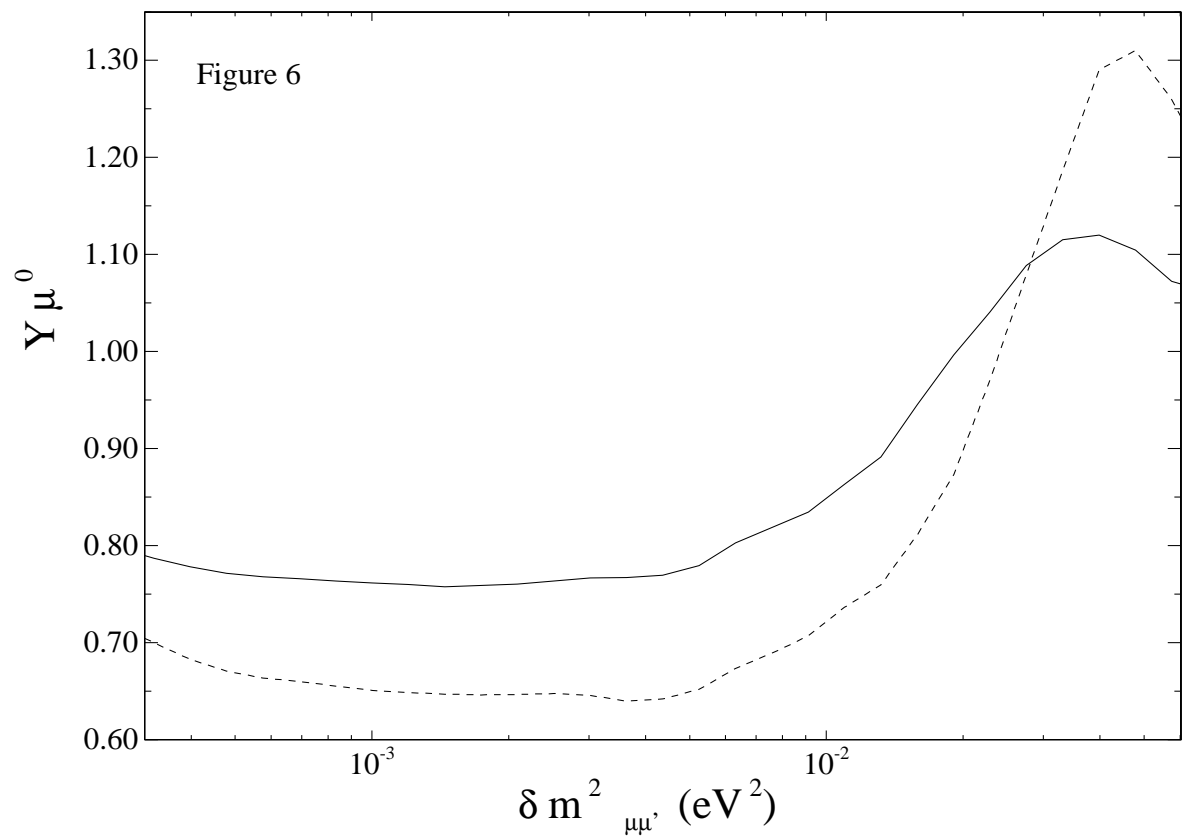
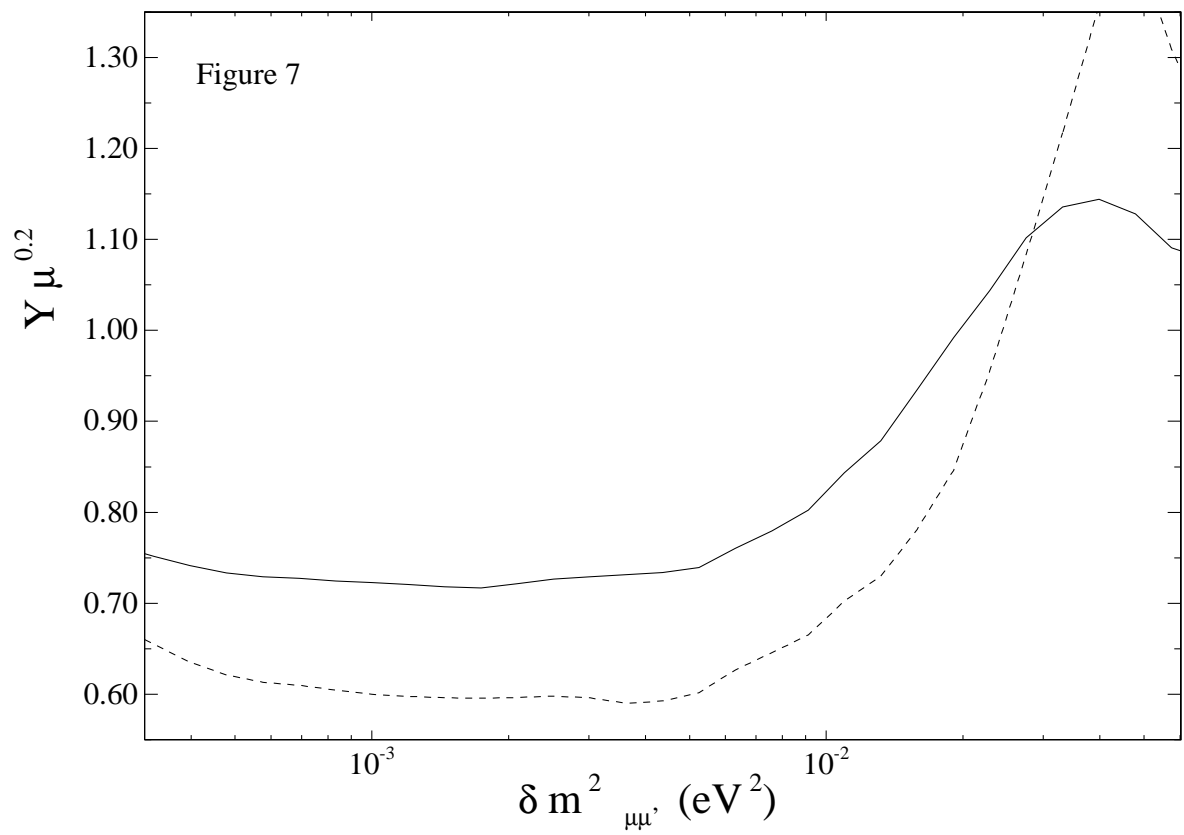
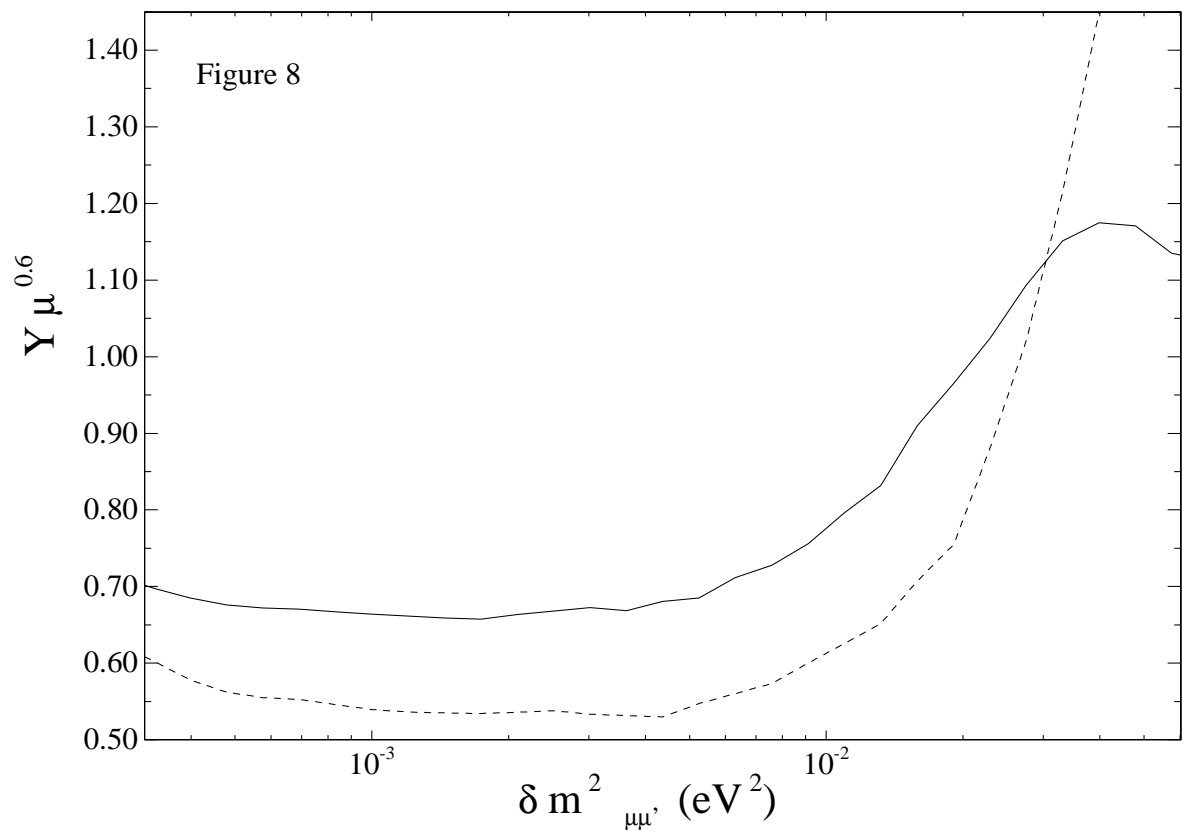
\end{document}